\documentstyle[eqsecnum,amssymb,prl,aps,multicol,psfig]{revtex}

\begin{document}
\draft
\title{Coulomb Explosion and Thermal Spikes}
\author{E. M. Bringa and R. E. Johnson}
\address{Engineering Physics University of Virginia Charlottesville VA 22903 U.S.A.}
\date{\today}
\maketitle
\begin{abstract}
A fast ion penetrating a solid creates a track of excitations. This can
produce displacements seen as an etched track, a process initially used to
detect energetic particles but now used to alter materials. From the seminal
papers by Fleischer et al. [1] to the present [2], `Coulomb explosion' and
thermal spike models are treated as conflicting models for describing ion
track effects. Here molecular dynamics simulations of electronic-sputtering,
a surface manifestation of ion track formation, show that `Coulomb
explosion' produces a `heat' spike so that these are early and late aspects
of the same process. Therefore, differences in scaling are due to the use of
incomplete spike models.
\end{abstract}

\pacs{PACS numbers:61.80.Az,79.20.-m,34.50.Fa,79.20.Ap}
\begin{multicols}{2}
\narrowtext

Since Thompson and Rutherford there has been interest in the track of
excitations produced by an energetic ion penetrating a solid. The fast ion
excites the electron cloud producing excitons and electron-hole pairs in an
insulator \cite{brown}. The electrons expelled from the track create
additional excitations, then cool to the lattice, screening and finally
recombining with the holes. They eventually return to the ground state by
radiative or non-radiative processes. Prior to this decay an excited region
with a net repulsive energy persists. Fleischer et al. \cite{fleischer}
(hereafter FPW) proposed that this repulsion could produce displacements
altering the material along the particle's track. When accounting only for
the holes the process has been called `Coulomb explosion', suggested to
produce amorphized tracks \cite{fleischer}, cratering, and sputtering \cite
{brown,haff}. Tracks have also been observed in liquids \cite{liquids},
semiconductors \cite{GaAs} and even metals \cite{dunlop} when the excitation
density in the track is sufficiently large.

In describing thresholds for track registration in insulators, FPW compared
their `Coulomb explosion' model to the often used thermal spike model for
defect production \cite{toul-Si,deformation}. Such a comparison is not only
of historical interest but persists in the recent literature \cite{trautmann}%
. However, one model describes a mechanism for energy-input (electrostatic
potential energy into kinetic energy), whereas the other describes the
transport of energy out of a `heated' cylindrical region. Here we present a
molecular dynamics (MD) description of the response of a solid to a
repulsive track. We show that at high excitation densities the primary
effect {\it is} the production of a cylindrical `heat' spike. Therefore,
distinctions made between `Coulomb explosion' and spike models can be
artifacts of the approximations used.

MD simulations are carried out for a model solid with a surface because of
our interest in electronically-induced sputtering, the surface manifestation
of track formation in insulators \cite{Johns}. As in track registration,
both spike models \cite{pospieszalska} and `Coulomb explosion' models \cite
{brown,tombrello} have been employed to parametrize laboratory data for
electronic sputtering. In spike models the atoms in the ion track are
assumed to have a radial temperature profile determined by the energy
deposited per unit path length, $(dE/dx)$. When the energy of an atom
exceeds a barrier, defect production in the bulk or evaporation at the
surface can occur. From the time dependent surface temperature, $T_{surf}$,
the spike sputtering yield, $Y_{S}$, the number of atoms ejected per ion
track is calculated integrating the surface flux over time and area. When a 
{\it radial diffusion} equation is used to determine $T_{surf}$, then $%
Y\,_{S}\propto \left( dE/dx\right) _{eff}^{2}$ for a fixed track radius at
high $(dE/dx)_{eff}$ \cite{mario,prb}, where $(dE/dx)_{eff}$ is the fraction
of $dE/dx$ going into energetic non-radiative processes \cite{brown}. This
has been used to describe sputtering produced by a track `heated' by
individual repulsive decay events \cite{pospieszalska}, secondary electrons 
\cite{trautmann} and a repulsive track \cite{brown,tombrellop,stampfli}.
However, when conditions leading to the $\left( dE/dx\right) _{eff}^{2}$
dependence {\it should} apply, the model fails because the energy transport
is {\it not }diffusive \cite{paperI}. A melt and a pressure pulse control
the energy transport leading to \cite{prb},

\begin{equation}
Y_{S}\approx 0.18(r_{cyl}/U)(dE/dx)_{eff}  \label{Ylinear}
\end{equation}
where $U$ is the solid's cohesive energy and the effect depth scales with $%
r_{cyl}$, the initial mean radius of the spike. Here we show that a {\it %
repulsively-induced} `heat' spike is produced in an ion track over a broad
range of excitation densities and the transport is {\it not} diffusive.
Therefore, a correct spike model must be used.

We simulate the evolution of a track of repulsive energy that might be
produced by a fast incident ion. Since sputtering \cite{prb} and crater
formation in both amorphous and crystalline, atomic or molecular solids can
be scaled over a broad range of material properties, we use a model solid
made of atoms interacting via Lennard-Jones (LJ) potentials. We use
parameters for condensed gas solids, but the results scale with the energy
and length parameters. Also, we showed earlier that scaling was maintained
using more complex potentials \cite{prb}. Interactions occur between all
atoms within a cut-off radius $r_{cut}\approx 2.54l$, where $l$ is the mean
atom spacing. The (001) layer spacing in the fcc lattice is $l_{s}\approx
0.8l$. MD simulations focusing on damage have been made for repulsive energy
between a few charges in LiF \cite{walkup}, a distribution of holes in Si 
\cite{cheng}, and using an assumed velocity distribution \cite{fedotov}.
Here we follow the conversion of a track of repulsive energy into atomic
motion in a solid.

On electronically exciting a condensed gas solid, the track of excitations
can produce large sputtering yields at high $dE/dx$, a process relevant to
icy outer solar system bodies\cite{Johns}. Ejection is presumed to be due to
the net repulsive energy, but a quantitative description is lacking. At high
ionization densities the net repulsion can persist if the holes have low
mobility and are not sufficiently screened during the time displacements
could occur. Even when holes are neutralized, the excited atoms have
overlapping charged clouds which act repulsively at high $dE/dx$ \cite
{brown,tombrellop}. Here we describe the net repulsion between `excited'
atoms in a track using $V=\left( e^{2}/r\right) \exp \left( -r/a\right) $,
where $a$ is an average screening constant \cite{nestor}. We do not
distinguish between closely spaced, partially screened holes or overlapping
excited species, since in both cases the electrons screen the interactions
between `excited' neighbors \cite{tombrellop,stampfli}. The interactions of
`excited' atoms with unexcited neighbors was left unchanged; including a
weak polarization had a small effect. For energy conservation, a larger
cut-off, $r_{cut}^{coul}=7a$, is used for interactions between `excited'
species.

\begin{figure}[htb]
\centerline{\psfig{figure=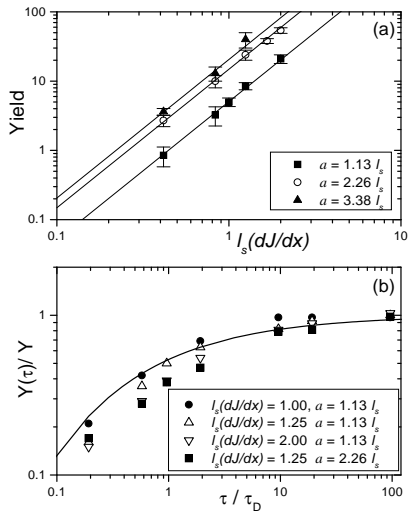,width=6.1cm,height=7.9cm,angle=0}}
\caption{(a). The sputtering yield from a repulsive
track vs. $\left[ l_{s}\left( dJ/dx\right) \right]$. The solid lines are %
$Y_{fit}=14.\,\ln \left[ 1.3\left( a/r_{coul}\right) \right]%
[l_{s}\,\,dJ/dx]^{2}$. [Results scale with LJ parameters. Those used here
are $\varepsilon =10.3$ meV and $\sigma =3.405$ \AA\ giving a cohesive
energy characteristic of a number of condensed-gas solids, $U=0.08$ eV and 
$l_{s}=2.66$ \AA .] (b). Yield scaled to the fit $Y_{fit}$ in Fig. 1a vs. 
$\tau $, the neutralization/recombination time. These are roughly fit by 
$exp[-\alpha (\tau _{D}/\tau )^{x}]$ with $\alpha $ and $x$ varying slowly
with $[l_{s}\,\,dJ/dx]$; curve shown is for $a=1.13l_{s}$, %
$[l_{s}\,dJ/dx]=1$ with $\alpha =0.46$ and $x=0.56$.
}
\label{fig1}
\end{figure}

The material is `excited' by instantaneously changing the potentials between
atoms in the track. The resulting velocities and positions of all particles
are then followed \cite{paperI}. This is done for a number of excitation
densities and screening constants. In a second set of calculations the
excitations were quenched (recombination). This was done statistically so
the average number of excitations in the track decayed exponentially, $\exp
\left[ -t/\tau \right] $ where $\tau $ is the quenching (neutralization)
time. The number of excitations per unit path length, $dJ/dx$, is related to 
$dE/dx$ for fast ions \cite{fleischer} and the mean radius of the
distributions, $r_{coul}$, depends on the speed of the incident ion. For
given track parameters [$dJ/dx$, $r_{coul}$ (here $\sim l_{s}$)] excited
species are chosen randomly. On excitation, atoms are sputtered if they
cross a plane $2r_{cut}^{coul}$ above the `top', here the (001) surface. The
number of atoms ejected in each run is called the yield, which is related to
the track of damage in FPW\ but is more easily measured. Sample depth was
chosen to be at least twice $r_{cut}^{coul}$. Different boundary conditions
did not change the average yields significantly. The sample size $%
\;(3x10^{4}-3x10^{5})$ and simulation times ($15-80$ ps) were adjusted to $a$
and $dJ/dx$. Extending times by tens of picoseconds or doubling the
thickness did not change the average yield. The distribution in the yield is
broad, especially for small $l_{s}(dJ/dx)$ where sputtering occurs when two
excitations are produced close together \cite{brown,haff}. Results, averaged
over $\sim 10-200$ excitation distributions, are given in Fig. 1a as a
function of charge density, $l_{s}(dJ/dx)$, for 3 values of the screening
constant $a$. The yield is seen to be {\it quadratic} in $dJ/dx$ for each $a$
and increases non-linearly with $a$; lines $Y=14.\ln \left[ 1.3\left(
a/r_{coul}\right) \right] \left[ l_{s}\left( dJ/dx\right) \right] ^{2}$ give
a good fit \cite{brown}.\ The yields as a function of neutralization time, $%
\tau ,$ for a fixed $(dJ/dx)$ and $a$ are given in Fig. 1b scaled by the fit
in Fig 1a. Because neutralization is treated stochastically, the size of the
yield is affected even for relatively large $\tau $ and only becomes
independent of the neutralization rate for $\tau \gtrsim 10\tau _{D}$, with $%
\tau _{D}$ the Debye period for the lattice ($\sim 0.5$ ps) \cite{paperI}.
However, the yield is still seen to be nearly quadratic in $(dJ/dx)$ for the
full range of $\tau $ with an additional weak dependence on excitation
density.

In MD simulations details of the energy transfer and sputtering can be
extracted. At very low excitation densities $(Y\lesssim 1)$ the probability
of neighbors being excited near the surface leads to ejection. At the
highest excitation densities $(Y\gtrsim 5)$, the repulsive energy is
transferred to neighbors in times $\sim 0.2\tau _{D}$ producing a
cylindrically heated region. Comparing the evolution of the resulting radial
temperature profiles with our earlier simulations of `heat' spikes, the
energy transport {\it is found to be similar}. That is, we find for the
repulsively produced heat spike transport is {\it not diffusive }\cite
{paperI}, a melt and a pressure pulse are formed and these control the
energy transport. Seiberling et al. \cite{tombrello} suggested that a
repulsive track could produce a heat spike but used the standard diffusive
model to obtaining a yield dependence very different from that calculated
here.

\begin{figure}[htb]
\centerline{\psfig{figure=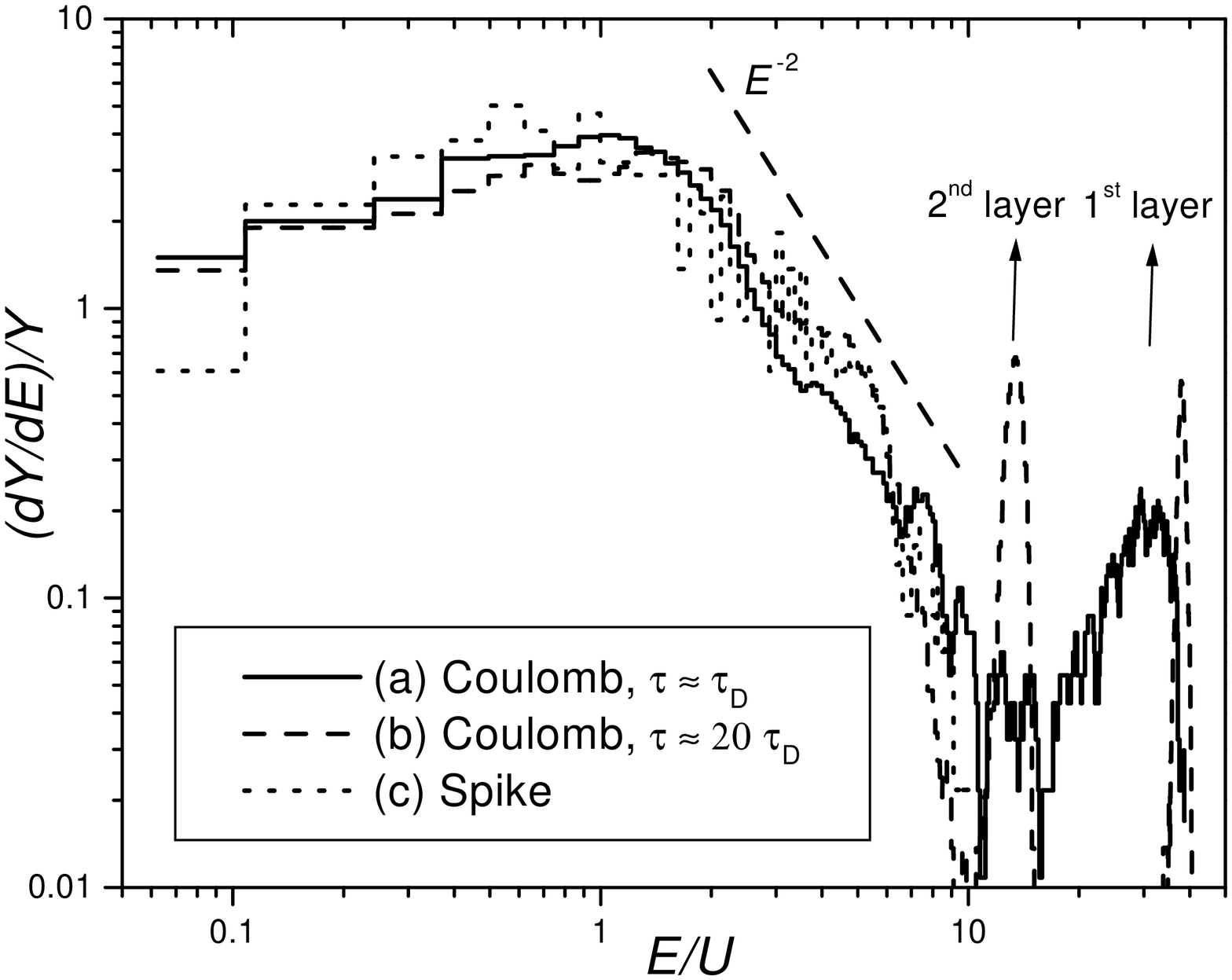,width=6.5cm,height=5.02cm,angle=0}}
\caption{%
Energy distribution of the ejecta for $\left[l_{s}\,dJ/dx\right] =2$, 
$a=1.13l_{s}$ and $r_{coul}=l_{s}$, for $\tau =\tau_{D}$ (a) and 
$\tau =20\tau _{D}$ (b). The dotted line shows the energy spectrum from 
a cylindrical spike, as in ref. \protect\cite{prb}, with 
$(dE/dx)_{eff}\approx 45U/l_{s}$ and $r_{cyl}\approx 2.6\ l_{s}$. For the
``Coulomb'' spike produced in (b), $(dE/dx)_{eff}\approx 38U/l_{s}$ and 
$r_{cyl}\approx 2.6\ l_{s}$, while the spike in (b) has a 30\% lower 
$(dE/dx)_{eff}$. Peaks are prompt ejecta, with position determined by the
potential energy between neighbors and the surface binding energy.}
\label{fig2}
\end{figure}

From the energy distributions for the ejecta at high excitation density two
regions can be seen in Fig. 2. At large ejecta energies, $E$ $\geq 10U$, the
spectra has peaks due to prompt ejection from the upper surface layers of
the initial track. This accounts for $\sim 20$\% of the ejecta at large $%
dJ/dx$ but dominates at very small $dJ/dx$. On the other hand, the principal
component of the ejecta in Fig. 2 exhibits an energy distribution like that
found in our studies of ejection from a narrow cylindrical heat spike shown
as the dashed line \cite{paperI}. That is, there is a broad, quasi-thermal
distribution at low ejecta energies, $E<U$ which gives way to a
non-Maxwellian, $\sim E^{-2}$ dependence at $E>U$. The latter dependence is
characteristic of low energy cascades in a solid \cite{prb} but differs from
thermal spike model predictions. Surprisingly, although the ejection process
changes in going from the lowest to the highest $dJ/dx$, there is no
dramatic change in the dependence of the yield on $dJ/dx$ in Fig. 1. This is
due to the nearly linear dependence of spike yield on $(dE/dx)_{eff}$
discussed below.

From the above, the primary effect of the repulsive decay at the high
excitation densities is the production of a `heat' spike. This is also shown
quantitatively using the MD calculation to determine the repulsive energy
driving atomic motion in times short compared to desorption times. At $\sim
0.2\tau _{D}$ the energy density deposited via repulsion is $%
(dE/dx)_{rep}\approx 0.15e^{2}(dJ/dx)^{2}$ for $a=1.13l_{s}\ $and $%
r_{coul}\approx l_{s}$. This energy is localized in a cylindrical region of
radius $r_{rep}\approx $ $2.6\ l_{s}$ for our values of $r_{coul}$. Using $%
(dE/dx)_{eff}=(dE/dx)_{rep}$ and $r_{cyl}=r_{rep}$ as initial conditions in
the expression for the yield from a cylindrical `heat' spike in Eq. \ref
{Ylinear} gives a yield that has {\it the same dependence on }$dJ/dx$ and a
size within 10\% of those in Fig.1a. Therefore, if a repulsive region is
sustained, a `heat spike' is formed at high $dJ/dx$ which determines the
subsequent energy transport, sputtering and displacements. Therefore,
`Coulomb explosion' and spikes are the {\it early and late }aspects of the
same process, so that contradictions for effects occurring later than $\sim
0.2\tau _{D}$ are due to incorrect descriptions of the `heat' spike. Below
we examine under what conditions a `Coulomb' track is sustained long enough
to produce a `heat' spike.

\begin{figure}[htb]
\centerline{\psfig{figure=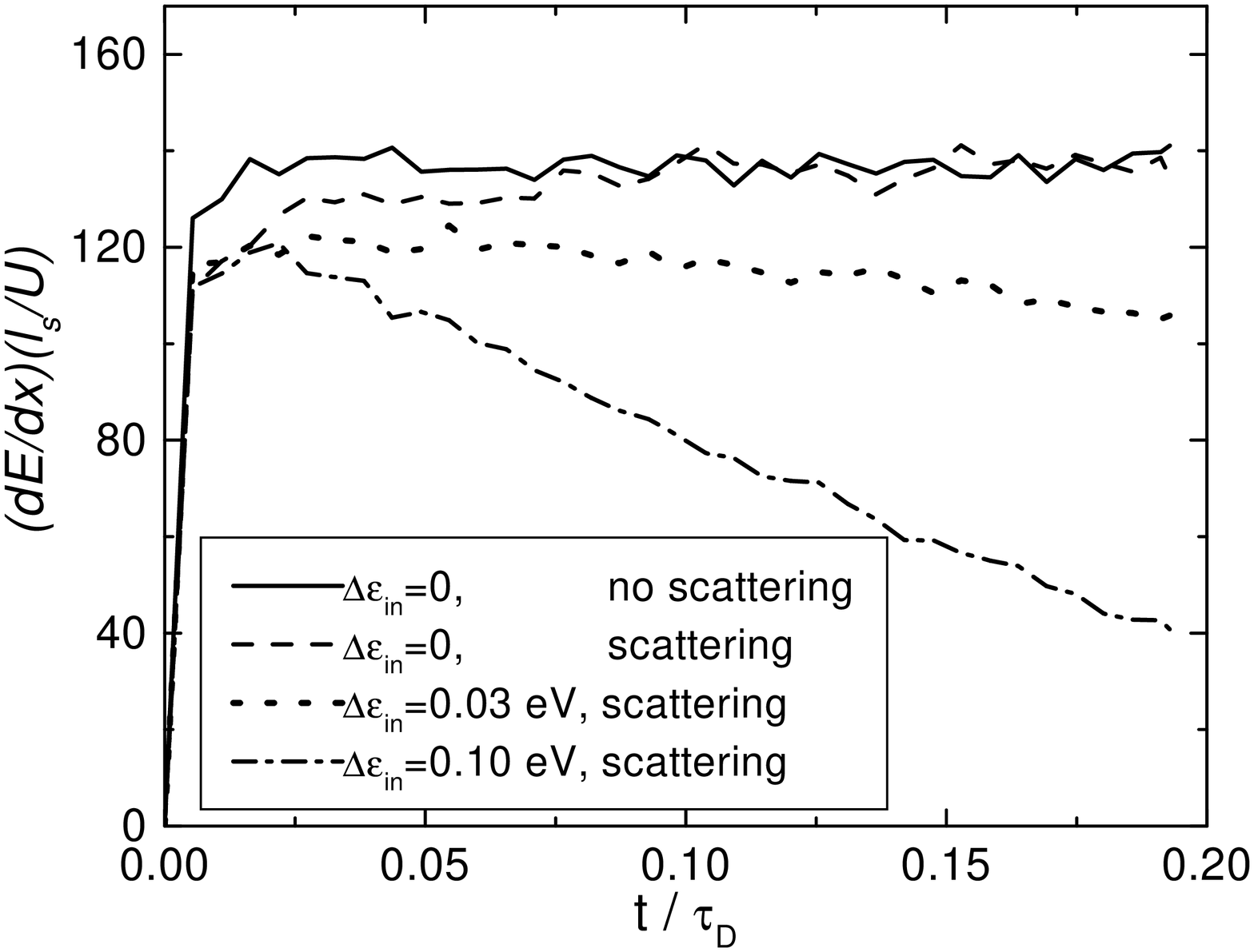,width=6.5cm,height=5.02cm,angle=0}}
\caption{The change in the total energy in the track
of interacting holes and electrons vs. time for `excited' electrons with $%
\left[ l_{s}\,dJ/dx\right] =1$, and a dielectric constant $\epsilon
=1.6\epsilon _{o}$. A fraction of the energy will be converted to atomic
motion. No scattering or inelastic energy loss (solid), no inelastic energy
loss but\thinspace scattering $\lambda _{s}=l$ (dashed), for $\lambda
_{in}=l $ and$\,\Delta \varepsilon _{in}=0.03$ eV (dotted), for$\,\,\lambda
_{in}=l$ and $\Delta \varepsilon _{in}=0.1$ eV (dash-dotted).}
\label{fig3}
\end{figure}

Since neutralization quenches the energy of a spike, we simulated the
neutralization and screening of holes by free electrons. That is, we make a
separate simulation in which we track the cooling of the `excited' electrons
in the field of the track of positive charges. The lattice is like that in
the MD simulations and we choose appropriate mean-free paths for elastic, $%
\lambda _{s}$, or inelastic, $\lambda _{in}$, scattering, both assumed to be
isotropic \cite{sanche}. A time dependent $a$ or an average $a$ and $\tau $
can be determined from such a simulation. Instead, we give in Fig. 3 the
instantaneous total potential energy density in the track vs. time. This is
the energy available for repulsive heating of the lattice. The interactions
between the positive charges in the track and the electrons are Coulombic
beyond an atomic radius, $a_{o}$. Inside $a_{o}$ the potentials gradually
become flat giving a binding energy of $\sim 8.6$ eV. We examined a number
of potential forms inside $a_{o}$ and initial electron energies and the
trends are the same. Electrons are `excited' by receiving, on the average,
15 eV, roughly the energy dissipated to the lattice (W-value minus
ionization energy\cite{sanche}). This does not treat the very fast electrons
which would further slow neutralization.

For only elastic scattering, the total energy in the track in Fig. 3 is
sustained with a very large effective $a$. Examining the radial dependence
of the electron density, a fraction of the electrons cool rapidly by
electron-electron collisions partially neutralizing the track. This is
dielectronic recombination, also called Auger recombination. The remainder
of the electron cloud, now `hotter' but more fully screened, expands to a 
{\it larger} average radius. Varying $\lambda _{s}$ affects the electron
density distribution, whereas including an inelastic energy loss, $\Delta
\varepsilon _{in}$, decreases the potential energy in the track. This also
heats the lattice, but here we are interested in the repulsive energy. For a 
$\Delta \varepsilon _{in}$ and $\lambda _{in}$ roughly corresponding to
electrons in a polymer or in ice (dotted curve) \cite{sanche}, it is seen in
Fig. 3 that most of the initial track energy remains during the time it
takes to produce a `heat spike' repulsively, $\sim 0.2\tau _{D}$. This would
correspond to an intermediate $\tau $ and relatively large $a$ in Figs. 1b
and 2a. Therefore, for a reasonable inelastic energy loss, screening is not
sufficient to quench the repulsive production of a `heat spike'. Increasing $%
\Delta \varepsilon _{in}$ (dot-dashed curve) or decreasing $\lambda _{in}$
leads to more rapid but still incomplete quenching over the relevant time.
Quantities for specific measurements need to be used and the role of excited
states in the track must be included.

In this paper we close the circle on one aspect of an old \cite{fleischer}
but topical \cite{trautmann} problem, the effect on a solid of a track of
excitations produced by a fast penetrating ion. A MD simulation of repulsive
explosion in a track of excitations has been carried out and the effect of
neutralization has been examined. This was done for narrow screened Coulomb
spikes in a condensed-gas solid to constrain the simulation size. However,
we showed earlier that the yields scale with $U$ and $(dE/dx)_{eff}$.
Whereas at low excitation densities ejection can occur if neighboring
`excited' species are formed near the surface \cite{brown,haff}, at the
higher excitation densities the repulsive energy in the track produces a
`heat' spike. Therefore, `Coulomb explosion' and spike models are the early
and late aspects of the repulsive decay process. The spike formed by the
repulsive heating of the lattice can lead to displacements and ejection
(electronic sputtering), examined here. Because the yield is roughly linear
in $(dE/dx)_{eff}$ at high excitation densities and $(dE/dx)_{eff}$ is
proportional to $(dJ/dx)^{2}$, the yield from a repulsive track is seen
(Fig.1a) to vary smoothly, having the same quadratic dependence on $dJ/dx$
in going from low to high excitation densities although the character of the
ejection process changes. At high excitation densities the yield can now be
estimated by substituting the repulsive energy deposition into the new,
useful expression for spike sputtering in Eq. \ref{Ylinear}. At higher
excitation densities than those studied here, the more distributed `heat'
spike produced by the electron cooling to the lattice might also be
important \cite{toul-Si}. Earlier applications of a `heat' spike for
describing track formation or electronic sputtering foundered on the use of
a spike model in which the energy transport was described incorrectly. That
is, the transport is not diffusive \cite{prb,paperI} and the local kinetic
energy distribution is not Maxwellian (Fig. 2). Calculations should now
focus on accurately describing the interactions in an ion track for times $%
\lesssim 0.2\ \tau _{D}$ including electrons, holes and excitons, as the
latter also contribute to the net repulsive energy.

\acknowledgments
We thank M. Liu for the neutralization calculations, H. Urbassek, R. Vidal,
and R. Baragiola for comments, the NSF Astronomy and Chemistry Divisions for
support.

\end{multicols}
\end{document}